\documentstyle{article}

 \selectfont 

\title{{\bf{ Nonlinear Schr\"{o}dinger equations and  N=2\\ superconformal algebra}}}
\author{\bf{H.\ T.\"Ozer }\thanks{ Corresponding author. \bf{E-mail\ :\ ozert @
itu.edu.tr }}
 \\\\
 Physics Department,\\ Faculty of Science and Letters,\\
Istanbul Technical University,\\
34469,\ Maslak,\ Istanbul,\ Turkey
}
\begin{document}
\maketitle
\par ~~~~~We obtain new coupled super Nonlinear
Schr\"odinger equations by using AKNS scheme and  soliton connection taking values in N=2
superconformal algebra .
\setcounter{equation}{0}

\vskip 5mm

\section{\bf{Introduction}}

\par~~~~~ Coupled  Nonlinear Schr\"odinger (NLS) equations can be obtained
 using  Ablowitz, Kaub,  Newell , Segur (AKNS) scheme[1].Extensions of
 coupled NLS equations have been obtained using a simple Lie algebra[2],
a Kac-Moody algebra [3],a Lie superalgebra[3,4],a Virasoro algebra[5] and
a N=1 superconformal algebra [6]in the literature.

\par ~~~~~The N=2 superconformal algebras were discovered in the seventies
independently by Ademollo et al.[7] and by Kac[8].The first authors derived
the algebras for physical purposes, in order to define supersymmetric strings,
whereas Kac derived them for mathematical purposes along with his classification
of Lie superalgebras.

\par ~~~~~The N=2 superconformal algebras provide the symmetries underlying
the N=2 strings[9,10,11,12,13]. These seem to be related to M-theory since
many of the basic objects of M-theory are realized in the heterotic (2,1) N=2
strings[14]. In addition, the topological version of the N=2 superconformal
algebra is realized in the world-sheet of the bosonic string[15],as well as in the
world-sheet of the superstrings[16].

\par ~~~~~In literature super-extensions of Nonlinear Schroedinger equations involve
finite number of bosonic and finite number of fermionic fields
[3,4,17,18,19].The super-extensions of Nonlinear Schroedinger
equations with infinite number of bosonic fields and infinite number
of fermionic fields has been studied in ref.6 and in this
paper.Mathematically this paper contains extension of the work done
in ref.6.

\par ~~~~~In  this paper
we will obtain super - extensions of coupled NLS equations using N=2
superconformal algebra with Neveu-Schwarz  and Ramond types . In
sec.2 we will discuss the sl(2,1)superalgebra valued soliton
connection and we will obtain coupled super NLS equations. Sec.3 and
sec.4 concern the soliton connection for the N=2 superconformal
algebra with Neveu-Schwarz type and Ramond type, respectively and we
will obtain in these sections two different types of super-
extensions of coupled NLS equations.

\newpage

\setcounter{equation}{0}
\section{\bf{ AKNS Scheme with sl(2,1) Superalgebra}}
\par  In  AKNS scheme in 1+1 dimension the  connection is defined as

\begin{equation}
\label{8}
\begin{array}{lll}
\Omega=\Omega_b+\Omega_f
\end{array}
\end{equation}
\noindent where
\begin{equation}
\label{8}
\begin{array}{lll}
\Omega_b= \Big( i \lambda J_1 +i \lambda H_1 +  Q^{+1} E_{+1}+Q^{-1}
E_{-1}  \Big) dx+
 \Big( -A_1 J_1-A_2 H_1 + B^{+1} E_{+1}+B^{-1} E_{-1}+ \Big) dt%
\end{array}
\end{equation}
\begin{equation}
\label{8}
\begin{array}{lll}
\Omega_f= \Big( P_1^{+{1\over2}} F_{+{1\over2}}+P_1^{-{1\over2}}
F_{-{1\over2}} + P_2^{+{1\over2}}
\bar{F}_{+{1\over2}}+P_2^{-{1\over2}} \bar{F}_{-{1\over2}}
 \Big) dx+
 \Big(
C_1^{+{1\over2}} F_{+{1\over2}}+C_1^{-{1\over2}} F_{-{1\over2}} +
C_2^{+{1\over2}} \bar{F}_{+{1\over2}}+C_2^{-{1\over2}}
\bar{F}_{-{1\over2}}
 \Big) dt%
\end{array}
\end{equation}
\noindent where  $J_1$,$H_1$, $E_{\pm1}$are bosonic generators and
$F_{\pm{1\over2}}$$\bar{F}_{\pm {1\over2}}$ are fermionic generators
of  $sl(1,2)\cong sl2,1)$ superalgebra which is the (N=2)extended
supersymmetric version of sl(2) algebra. These generators have
matrix representations as
\begin{equation}
\begin{array}{ccc}
H_1 = \left(\begin{array}{cccc}
                               {1\over2} & 0 & 0 \\
                               0 &-{1\over2} & 0 \\
                               0 & 0 & 0 \end{array}
                        \right);
J_1 = \left(\begin{array}{cccc}
                               {1\over2} & 0 & 0 \\
                               0 &{1\over2} & 0 \\
                               0 & 0 & 0 \end{array}
                        \right);
E_{+1} = \left(\begin{array}{cccc}
                               0 & 1 & 0\\
                               0 & 0 & 0\\
                               0 & 0 & 0 \end{array}
                        \right);
E_{-1} = \left(\begin{array}{cccc}
                               0 & 0 & 0\\
                               1 & 0 & 0\\
                               0 & 0 & 0 \end{array}
                        \right)\\
F_{+{1\over2}}= \left(\begin{array}{cccc}
                               0 & 0 & 0 \\
                               0 & 0 & 0 \\
                               0 & 1 & 0 \end{array}
                        \right);
F_{-{1\over2}} = \left(\begin{array}{cccc}
                               0 & 0 & 0 \\
                               0 & 0 & 0 \\
                               1 & 0 & 0 \end{array}
                        \right)

\bar{F}_{+{1\over2}}= \left(\begin{array}{cccc}
                               0 & 0 & 1 \\
                               0 & 0 & 0 \\
                               0 & 0 & 0 \end{array}
                        \right);
\bar{F}_{-{1\over2}} = \left(\begin{array}{cccc}
                               0 & 0 & 0 \\
                               0 & 0 & 1 \\
                               0 & 0 & 0 \end{array}
                        \right)
\end{array}
\end{equation}
\noindent Also, these generators satisfy the following commutation
and anticommutation  relations

\begin{equation}
\label{17}
\begin{array}{ccc}
\begin{array}{cccc}

 \left[ H_1,E_{\pm 1 }\right]  = \pm  E_{\pm 1};&
 \left[ H_1,F_{\pm 1 }\right]  = \pm {1\over2} F_{\pm 1}  ;&
 \left[ H_1,\bar{F}_{\pm 1 }\right] = \pm {1\over2} \bar{F}_{\pm
1}\\
 \left[ J_1,H_{\pm 1 }\right]  =\left[ J_1,E_{\pm 1 }\right]=0  ;&
\left[ J_1,F_{\pm 1 }\right]  =  {1\over2} F_{\pm 1} ;&
\left[ J_1,\bar{F}_{\pm 1 }\right]  =  -{1\over2} \bar{F}_{\pm 1}  \\

\left[E_{\pm 1 } ,F_{\mp {1\over2}}\right]  = \left[E_{\pm 1 }
,\bar{F}_{\mp {1\over2}}\right]=0;&\left[E_{\pm 1 },F_{\mp
{1\over2}}\right]  =-F_{\pm {1\over2}};& \left[E_{\pm 1
},\bar{F}_{\mp {1\over2}}\right]  =\bar{F}_{\pm
{1\over2}}\\

\left\{F_{\pm 1 },F_{\pm {1\over2}}\right\}  =\left\{\bar{F}_{\pm 1
},\bar{F}_{\pm {1\over2}}\right\}=0;& \left\{F_{\pm 1 },F_{\mp
{1\over2}}\right\}  =\left\{\bar{F}_{\pm 1 },\bar{F}_{\mp
{1\over2}}\right\}=0 ;&\left\{F_{\pm 1 },\bar{F}_{\mp
{1\over2}}\right\}  =E_{\pm 1 }\\

   \left[E_{+1},E_{-1}\right]  =  H_1 ;&
    \left\{F_{\pm {1\over2}} ,\bar{F}
    _{\pm {1\over2}}\right\}  =  J_1\mp H_1\\
\end{array}
\end{array}
\end{equation}
\noindent In Eq.(1) $\lambda$ is the spectral parameter, $Q^{\pm
1}$,$P_1^{\pm {1\over2}}$ and $P_2^{\pm {1\over2}}$  are fields
depending on space and time, namely x and t, and functions
$A_1$,$A_2$,$B^{\pm 1}$,$C_1^{\pm {1\over2}}$ and $C_2^{\pm
{1\over2}}$ are x,t and $\lambda$ dependent.
The integrability condition is given by
\begin{equation}
\label{8}
\begin{array}{lll}
d\ \Omega\  + \ \Omega\ \wedge\ \Omega  =  0
\end{array}
\end{equation}
\par By using Eqs.(1) and (6) one can obtain following equations:
\begin{equation}
\label{19}
\begin{array}{lll}
A_{1x}
=P_1^{-{1\over2}}C_2^{+{1\over2}}
+P_1^{+{1\over2}}C_2^{-{1\over2}}
+P_2^{-{1\over2}}C_1^{+{1\over2}}
+P_2^{+{1\over2}}C_1^{-{1\over2}}
\end{array}
\end{equation}
\noindent
\begin{equation}
\label{19}
\begin{array}{lll}
A_{2x}=
- 2 Q^{-1}B^{+1}
+ 2 Q^{+1}B^{-1}
+P_1^{-{1\over2}}C_2^{+{1\over2}}
-P_1^{+{1\over2}}C_2^{-{1\over2}}
-P_2^{-{1\over2}}C_1^{+{1\over2}}
+P_2^{+{1\over2}}C_1^{-{1\over2}}
\end{array}
\end{equation}
\noindent
\begin{equation}
\label{18}
\begin{array}{lll}
{Q^{+ 1}}_t=\ \ {B^{+ 1}}_x+  i \lambda  B^{+1} +  Q^{+1} A_2
+P_1^{+{1\over2}}C_2^{+{1\over2}}+P_2^{+{1\over2}}C_1^{+{1\over2}}
\end{array}
\end{equation}
\noindent
\begin{equation}
\label{19}
\begin{array}{lll}
{Q^{- 1}}_t=\ \ {B^{- 1}}_x-  i \lambda  B^{-1} -  Q^{-1} A_2
+P_1^{-{1\over2}}C_2^{-{1\over2}}+P_2^{-{1\over2}}C_1^{-{1\over2}}
\end{array}
\end{equation}
\noindent
\begin{equation}
\label{19}
\begin{array}{lll}
{P_1^{+{1\over2}}}_{t}=\ \ {C_1^{+{1\over2}}}_x
+{i} \lambda C_1^{+{1\over2}}
+P_1^{-{1\over2}}B^{+1}
+{1\over2} P_1^{+{1\over2}}A_1
+{1\over2} P_1^{+{1\over2}}A_2
-Q^{+1} C_1^{-{1\over2}}
\end{array}
\end{equation}
\noindent
\begin{equation}
\label{19}
\begin{array}{lll}
{P_1^{-{1\over2}}}_{t}=\ \ {C_1^{-{1\over2}}}_x
+P_1^{+{1\over2}}B^{-1}
+{1\over2} P_1^{-{1\over2}}A_1
-{1\over2} P_1^{-{1\over2}}A_2
-Q^{-1} C_1^{+{1\over2}}
\end{array}
\end{equation}
\noindent
\begin{equation}
\label{19}
\begin{array}{lll}
{P_2^{+{1\over2}}}_{t}=\ \ {C_2^{+{1\over2}}}_x
-P_2^{-{1\over2}}B^{+1}
-{1\over2} P_2^{+{1\over2}}A_1
+{1\over2} P_2^{+{1\over2}}A_2
+Q^{+1} C_2^{-{1\over2}}
\end{array}
\end{equation}
\noindent
\begin{equation}
\label{19}
\begin{array}{lll}
{P_2^{-{1\over2}}}_{t}=\ \ {C_2^{-{1\over2}}}_x
-{i} \lambda C_2^{-{1\over2}}
-P_2^{+{1\over2}}B^{-1}
-{1\over2} P_2^{-{1\over2}}A_1
-{1\over2} P_2^{-{1\over2}}A_2
+Q^{-1} C_2^{+{1\over2}}
\end{array}
\end{equation}
\noindent
\par In AKNS scheme we expand $A_1$,$A_2$,$B^{\pm 1}$ $C_1^{\pm
{1\over2}}$ and $C_2^{\pm {1\over2}}$
in terms of  positive powers of $\lambda$ as
\begin{equation}
A_1=\sum_{\scriptstyle n=0}^2 \lambda^n a_{1n};\ \
A_2=\sum_{\scriptstyle n=0}^2 \lambda^n a_{2n};\ \ B^{\pm
1}=\sum_{\scriptstyle n=0}^2 \lambda^n b^{\pm 1}_n;\ \
C_1^{\pm {1\over2}}=\sum_{\scriptstyle n=0}^2 \lambda^n c^{\pm{1\over2}}_{1n};\ \
C_2^{\pm {1\over2}}=\sum_{\scriptstyle n=0}^2 \lambda^n c^{\pm{1\over2}}_{2n}
\end{equation}
\noindent Inserting Eq.(15) into Eqs.(7-14)gives 24 relations  in
terms of $a_{1n}$,$a_{2n}$,$b^{\pm 1}_n$,$c^{\pm {1\over2}}_{1n}$ and $c^{\pm {1\over2}}_{2n}$.
By solving these relations we get
$$a_{10}=
-i P_1^{-{1\over2}}P_2^{+{1\over2}}
-i P_1^{+{1\over2}}P_2^{-{1\over2}};\ \ a_{11}=0;\ \ a_{12}=-i;$$
$$a_{20}=
-2 i Q^{+1}Q^{-1}
-i P_1^{-{1\over2}}P_2^{+{1\over2}}
+i P_1^{+{1\over2}}P_2^{-{1\over2}};\ \ a_{21}=0;\ \ a_{22}=-i;$$
$$
b^{\pm 1}_0=\pm i {Q^{\pm 1}}_x+i P_2^{\pm {1\over2}}P_1^{\pm
{1\over2}}  ;\
 b^{\pm1}_1= Q^{\pm1};\ b^{\pm1}_2=0
 \eqno(16)
$$
$$
c^{\pm{1\over2}}_{10}=i {P_1^{\mp{1\over2}}}Q^{\pm1} + i{P_1^{\pm}}_x;\
c^{\pm{1\over2}}_{20}=i {P_2^{\mp{1\over2}}}Q^{\pm1} - i{P_2^{\pm}}_x
$$
$$
 c^{+{1\over2}}_{11}= P^{+{1\over2}};\
 c^{-{1\over2}}_{21}= P^{-{1\over2}};\
 c^{-{1\over2}}_{11}=c^{+{1\over2}}_{21}=0;\
 c^{\pm{1\over2}}_{12}=c^{\pm{1\over2}}_{22}=0
$$
\noindent By using the relations given by Eq.(16) from Eqs.(9-14) we
obtain the coupled super NLS equations as \setcounter{equation}{11}
\setcounter{equation}{16}
\begin{equation}
\label{118}
\begin{array}{lll}
{-i Q^{+1}}_t= {Q^{+1 }}_{xx}
-2 (Q^{+1 })^2 Q^{-1}
+2Q^{+1} P_2^{+{1\over2}}P_1^{-{1\over2}}
-2Q^{+1} P_2^{-{1\over2}}P_1^{+{1\over2}}
+2( P_2^{+{1\over2}}P_1^{+{1\over2}})_x\\
{~~~i Q^{-1}}_t= {Q^{+1 }}_{xx}
-2 (Q^{-1 })^2 Q^{+1}
-2Q^{-1} P_2^{-{1\over2}}P_1^{+{1\over2}}
+2Q^{-1} P_2^{+{1\over2}}P_1^{-{1\over2}}
-2( P_2^{-{1\over2}}P_1^{-{1\over2}})_x\\
{-i P_1^{+{1\over2}}}_t=
   {P_1^{+{1\over2} }}_{xx}
-2 P_1^{+{1\over2}} Q^{+1}Q^{-1}
+2 P_1^{-{1\over2}} {Q^{+1}}_x
-2 P_2^{+{1\over2}}P_1^{+{1\over2}} P_1^{-{1\over2}}\\
{-i P_1^{-{1\over2}}}_t=
   {P_1^{-{1\over2} }}_{xx}
-2 P_1^{-{1\over2}} Q^{+1}Q^{-1}
-2 P_2^{-{1\over2}}P_1^{-{1\over2}} P_1^{+{1\over2}}\\
{~~~i P_2^{+{1\over2}}}_t=
   {P_2^{+{1\over2} }}_{xx}
+2 P_2^{+{1\over2}} Q^{+1}Q^{-1}
+2 P_2^{+{1\over2}}P_2^{-{1\over2}} P_1^{+{1\over2}}\\
{~~~i P_2^{-{1\over2}}}_t=
   {P_2^{-{1\over2} }}_{xx}
-2 P_2^{-{1\over2}} Q^{+1}Q^{-1}
-2 P_2^{+{1\over2}} {Q^{-1}}_x
+2 P_2^{+{1\over2}}P_2^{-{1\over2}} P_1^{-{1\over2}}\\
\end{array}
\end{equation}
\section{\bf{AKNS Scheme with N=2 Superconformal Algebra (Neveu-Schwarz Type)}}
\par We generalize the connection given by Eq.(1) as
\setcounter{equation}{17}
\begin{equation}
\label{8}
\begin{array}{lll}
\Omega=\Omega_b+\Omega_f
\end{array}
\end{equation}
where
\begin{equation}
\label{8}
\begin{array}{lll}
\Omega_b=
& \Big(&
  i \lambda J_0
+ i \lambda L_0
+ Q_1^{+m} J_{+m}
+ Q_1^{-m} J_{-m}
+ Q_2^{+m} L_{+m}
+ Q_2^{-m} L_{-m}
 \Big) dx+\\
& \Big(&
-A_1 J_0
-A_2 L_0
+ B_1^{+m} J_{+m}
+ B_1^{-m} J_{-m}
+ B_2^{+m} L_{+m}
+ B_2^{-m} L_{-m}
 \Big) dt%
\end{array}
\end{equation}
\begin{equation}
\label{8}
\begin{array}{lll}
\Omega_f=
& \Big(&
+ P_1^{+{m\over2}} G^1_{+{m\over2}}
+ P_1^{-{m\over2}} G^1_{-{m\over2}}
+ P_2^{+{m\over2}} G^2_{+{m\over2}}
+ P_2^{-{m\over2}} G^2_{-{m\over2}}
 \Big) dx+\\
& \Big(&
+ C_1^{+{m\over2}} G^1_{+{m\over2}}
+ C_1^{-{m\over2}} G^1_{-{m\over2}}
+ C_2^{+{m\over2}} G^2_{+{m\over2}}
+ C_2^{-{m\over2}} G^2_{-{m\over2}}
 \Big) dt%
\end{array}
\end{equation}
\noindent where $L_0$ , $J_{\pm m}$ , $L_{\pm m}$ are bosonic generators and
$G^1_{\pm {m\over2}}$,$G^2_{\pm {m\over2}}$ are fermionic generators of centerless N=2
superconformal algebra of Neveu-Schwarz type .In modern notation namely this algebra  satisfy
the following commutation and anticommutation  relations
\begin{equation}
\label{34}
\begin{array}{lll}
\left[ L_r,L_s\right]&  = & (r-s)\ L_{r+s}\\
\left[ J_r,J_s\right]&  = & 0\\
\left[ L_r,J_s\right]&  = & -s\ J_{r+s}\\
\left\{ G^1_r,G^2_s\right\}&  = & 2\ L_{r+s}+ (r-s)\ J_{r+s}\\
\left[ J_r,G^{1,2}_s\right] & = & \pm G^{1,2}_{r+s}\\
\left[ L_r,G^{1,2}_s\right] & = & ({r\over2}-s)\ G^{1,2}_{r+s}\\
\left\{ G^{1,2}_r,G^{1,2}_s\right\}&  = & 0%
\end{array}
\end{equation}
\noindent  Here, $J_{\pm m}$,$L_{\pm m}$ are generators with
positive(negative) integer indices and $G^{1,2}_{\pm {m\over2}}$ are
generators with positive(negative) half integer indices. In
Eq.(18) we assume summation over the repeated indices. The fields
$Q_{1,2}^{\pm m}$ and
 $P_{1,2}^{\pm {m\over2}}$are x,t dependent and also functions $A_{1,2}$,
 $B_{1,2}^{\pm m}$ and $C_{1,2}^{\pm {m\over2}}$ are x,t and $\lambda$ dependent.
 \vskip 5mm
In N=2 superconformal algebra if we restrict $J_{\pm m}$ to have
only $J_0$ components,$L_{\pm m}$ to have only $L_0$ , $L_{\pm 1}$
components, and $G^{1,2}_{\pm {m\over2}}$ to have only $G^{1,2}_{\pm
{1\over2}}$ components we obtain sl(1,2) algebra given by Eq.(5)
with the following definitions:
\begin{equation}
\label{342}
\begin{array}{lll}
J_1 ={1\over2} J_0 ;\ \ H_1 =- L_0 ;\ \ E_{+1} =L_{+1} ;\ \ E_{-1} = -L_{-1}\\
 F_{+{1\over2}} =G^1_{+{1\over2}} ;\ \ F_{-} =
 -{1\over2}G^1_{-{1\over2}};
\bar{F}_{+{1\over2}} ={1\over2}G^2_{+{1\over2}} ;\ \
\bar{F}_{{1\over2}} = -{1\over2}G^2_{-{1\over2}}
\end{array}
\end{equation}
\noindent From the integrability condition given by Eq.(6) we obtain
$$
A_{1x}= \sum_{\scriptstyle r=1}^\infty
   r ( B_1^{-r} Q_2^{+r}
   - B_1^{+r} Q_2^{-r}
   + B_2^{-r} Q_1^{+r}
   - B_2^{+r} Q_1^{-r}
   - P_1^{-{r\over2}} C_2^{+{r\over2}}
   + P_1^{+{r\over2}} C_2^{-{r\over2}}
   + P_2^{-{r\over2}} C_1^{+{r\over2}}
   - P_2^{+{r\over2}} C_1^{-{r\over2}})\eqno(23)
$$
$$
A_{2x}=2 \sum_{\scriptstyle r=1}^\infty
   (r B_2^{-r} Q_2^{+r}
   -r B_2^{+r} Q_2^{-r}
   + P_1^{-{r\over2}} C_2^{+{r\over2}}
   + P_1^{+{r\over2}} C_2^{-{r\over2}}
   + P_2^{-{r\over2}} C_1^{+{r\over2}}
   + P_2^{+{r\over2}} C_1^{-{r\over2}})\eqno(24)
$$
\setcounter{equation}{24}
\begin{equation}
\label{8}
\begin{array}{lll}
{Q_1^{\pm m}}_t= {B_1^{\pm m}}_x \mp i m  \lambda B_1^{\pm m}\mp m
A_2 Q_1^{\pm m}+\delta_J^{\pm m}
\end{array}
\end{equation}
\begin{equation}
\label{8}
\begin{array}{lll}
{Q_2^{\pm m}}_t= {B_2^{\pm m}}_x \mp i m  \lambda B_2^{\pm m}\mp m A_2 Q_2^{\pm m}+\delta_L^{\pm m}
\end{array}
\end{equation}
\begin{equation}
\label{8}
\begin{array}{lll}
{P_1^{\pm {m\over2}}}_t= {C_1^{\pm {m\over2}}}_x\mp{i\over2}(m\mp 2) \lambda
C_1^{\pm {m\over2}}\mp {m\over2} P_1^{\pm {m\over2}} A_2+ P_1^{\pm {m\over2}}
A_1+\delta_{G^{1}}^{\ pm m}
\end{array}
\end{equation}
\begin{equation}
\label{8}
\begin{array}{lll}
{P_2^{\pm {m\over2}}}_t= {C_2^{\pm {m\over2}}}_x\mp{i\over2}(m\pm 2) \lambda
C_2^{\pm {m\over2}}\mp {m\over2} P_2^{\pm {m\over2}} A_2- P_2^{\pm {m\over2}}
A_1+\delta_{G^{2}}^{\pm m}
\end{array}
\end{equation}
\noindent where

$$
\delta_J^{+m}=\sum_{\scriptstyle r,s=1}^\infty \Big[(r
B_2^{+s}Q_1^{+r}- s B_1^{+s}Q_2^{+r} ) \delta_{r+s,m}
+(r-s)(P_1^{+{r\over2}} C_2^{+{s\over2}}-P_2^{+{r\over2}}
C_1^{+{s\over2}}) \delta_{r+s,2 m} \Big]
$$
$$
+\sum_{\scriptstyle r,s=1\atop\scriptstyle r>s}^\infty \Big[(r
B_2^{-s}Q_1^{+r}+ s B_1^{-s}Q_2^{+r} ) \delta_{r-s,m}
+{(r+s)\over2}(P_1^{+{r\over2}} C_2^{-{s\over2}}-P_2^{+{r\over2}}
C_1^{-{s\over2}}) \delta_{r-s,2 m} \Big]\eqno(29)
$$
$$
-\sum_{\scriptstyle r,s=1\atop\scriptstyle r<s}^\infty \Big[(r
B_2^{+s}Q_1^{-r}+ s B_1^{+s}Q_2^{-r} ) \delta_{-r+s,m}
+{(r+s)\over2}(P_1^{-{r\over2}} C_2^{+{s\over2}}-P_2^{-{r\over2}}
C_1^{+{s\over2}}) \delta_{-r+s,2 m} \Big]
$$
$$
\delta_L^{+m}=\sum_{\scriptstyle r,s=1}^\infty \Big[(r-s)
B_2^{+s}Q_2^{+r} \delta_{r+s,m} +2(P_1^{+{r\over2}}
C_2^{+{s\over2}}+P_2^{+{r\over2}} C_1^{+{s\over2}}) \delta_{r+s,2 m}
\Big]
$$
$$
+\sum_{\scriptstyle r,s=1\atop\scriptstyle r>s}^\infty \Big[(r+s)
B_2^{-s}Q_2^{+r} \delta_{r-s,m} +2(P_1^{+{r\over2}}
C_2^{-{s\over2}}+P_2^{+{r\over2}} C_1^{-{s\over2}}) \delta_{r-s,2 m}
\Big]\eqno(30)
$$
$$
-\sum_{\scriptstyle r,s=1\atop\scriptstyle r<s}^\infty \Big[(r+s)
B_2^{+s}Q_2^{-r} \delta_{-r+s,m} -2(P_1^{-{r\over2}}
C_2^{+{s\over2}}+P_2^{-{r\over2}} C_1^{+{s\over2}}) \delta_{-r+s,2
m} \Big]
$$
$$
\delta_{G^{1}}^{+m}=-{1\over2}\sum_{\scriptstyle r,s=1}^\infty
\Big[2 B_1^{+s}-(r-s)B_2^{+s})P_1^{+{r\over2}} \delta_{r+2 s,m} -2
(Q_1^{+r}+(r-s) Q_2^{+r})C_1^{+{s\over2}} \delta_{2r+s,m} \Big]
$$
$$
+{1\over2}\sum_{\scriptstyle r,s=1\atop\scriptstyle 2r>s}^\infty
\Big[2 Q_1^{+r}+(r+s)Q_2^{+r})C_1^{-{s\over2}} \delta_{2r-s,m}\Big]
-{1\over2}\sum_{\scriptstyle r,s=1\atop\scriptstyle r>2s}^\infty
\Big[2 (B_1^{-s}-(r+s) B_2^{-s})P_1^{+{r\over2}} \delta_{r-2s,m}
\Big]\eqno(31)
$$
$$
-{1\over2}\sum_{\scriptstyle r,s=1\atop\scriptstyle r<2s}^\infty
\Big[2 B_1^{+s}+(r+s)B_2^{+s})P_1^{-{r\over2}} \delta_{-r+2s,m}\Big]
+{1\over2}\sum_{\scriptstyle r,s=1\atop\scriptstyle 2r<s}^\infty
\Big[2 (Q_1^{-r}-(r+s) Q_2^{-r})C_1^{+{s\over2}} \delta_{-2r+s,m}
\Big]
$$
$$
\delta_{G^{2}}^{+m}=\delta_{G^{1}}^{+m}\left(%
\begin{array}{c}
  P_1\rightarrow -P_2 \\
  C_1\rightarrow-C_2 \\
  \end{array}%
\right)\eqno(32)
$$
\noindent and
$$
\delta_J^{-m},\\ \delta_L^{-m},\\ \delta_{G^{1}}^{-m},\\ \delta_{G^{2}}^{-m}
=\\ \delta_{G^{1}}^{+m},\\ \delta_{G^{2}}^{+m},\\ \delta_J^{+m},\\ \delta_L^{+m}\left(%
\begin{array}{c}
  +m\rightarrow-m \\
  +r\rightarrow-r \\
  +s\rightarrow-s \\
\end{array}%
\right)\eqno(33)
$$
\par In AKNS scheme we expand
$A_1$,$A_2$,$B^{\pm 1}$,$C_1^{\pm {1\over2}}$ and $C_2^{\pm
{1\over2}}$
 in terms of the positive powers of $\lambda$ as
\setcounter{equation}{33}
\begin{equation}
A_1=\sum_{\scriptstyle n=0}^2 \lambda^n a_{1n};\ \
A_2=\sum_{\scriptstyle n=0}^2 \lambda^n a_{2n}
;\ \ B_1^{\pm m}=\sum_{\scriptstyle n=0}^2 \lambda^n b^{\pm m}_{1n};\ \ B_2^{\pm
m}=\sum_{\scriptstyle n=0}^2 \lambda^n b^{\pm m}_{2n};\ \
\end{equation}
\begin{equation}
C_1^{\pm {m\over2}}=\sum_{\scriptstyle n=0}^2 \lambda^n c^{\pm{m\over2}}_{1n};\ \
C_2^{\pm {m\over2}}=\sum_{\scriptstyle n=0}^2 \lambda^n c^{\pm{m\over2}}_{2n}
\end{equation}
\noindent Inserting Eq.(34-35) into Eqs.(23-28)gives 30 relations  in
terms of $a_{1n}$,$a_{2n}$,$b^{\pm m}_{1n}$,$b^{\pm m}_{2n}$,$c^{\pm {m\over2}}_{1n}$ and
$c^{\pm {m\over2}}_{2n}$ (n=0,1,2) .
By solving these relations we get
$$
a_{10}= i \sum_{\scriptstyle r=1}^\infty
( Q_2^{+r}Q_1^{-r}
+ Q_2^{-r}Q_1^{+r})
-i \sum_{\scriptstyle  r=1}^\infty
 \Big[{{2}\over{r-2}}\Big]P_2^{-{r\over2}}P_1^{+{r}}
-i\sum_{\scriptstyle  r=1}^\infty
 \Big[ {{2}\over{r+2}}\Big]P_2^{+{r}}P_1^{-{r}}
;\ a_{11}=a_{12}=0;\ \
$$
$$
a_{20}= 2i \sum_{\scriptstyle r=1}^\infty
 Q_2^{+r}Q_2^{-r}
-i \sum_{\scriptstyle  r=1}^\infty
 \Big[{{4}\over{r-2}}\Big]P_2^{-{r}}P_1^{+{r}}
+i\sum_{\scriptstyle  r=1}^\infty
 \Big[ {{4}\over{r+2}}\Big]P_2^{+{r}}P_1^{-{r}}
;\ a_{21}=a_{22}=0;\ \
$$
$$
b^{\pm m}_{10}=\mp {i\over m} {Q_1^{\pm m}}_x ;\
b^{\pm m}_{20}=\mp {i\over m} {Q_2^{\pm m}}_x ;\
 b^{\pm m}_{11}= Q_1^{\pm m};\
 b^{\pm m}_{21}= Q_2^{\pm m};\ b^{\pm m}_{12}=0;\ b^{\pm m}_{22}=0
 \eqno(36)
$$
$$
c^{\pm {m\over2}}_{10}= \mp {{2i}\over {m-2}}{P_1^{\pm {m\over2} }}_x  ;\
c^{\pm {m\over2}}_{20}= \pm {{2i}\over {m+2}}{P_2^{\mp {m\over2} }}_x  ;\
c^{\pm {m\over2}}_{11}= P_1^{\pm {m\over2}};\
c^{\pm {m\over2}}_{21}= P_2^{\pm {m\over2}};\
c^{\pm {m\over2}}_{12}=c^{\pm {m\over2}}_{22}=0
$$
\noindent By using the relations given by Eq.(36) from Eqs.(25-28)
we obtain the coupled super NLS equations as
$$
{-i Q_1^{\pm m}}_t=\mp{{1}\over{m}}{Q_1^{\pm m}}_{xx}
-2   Q_1^{\pm m}\left(\sum_{\scriptstyle r=1}^\infty Q_2^{-r}Q_2^{+r}\right)
 \pm 4  m Q_1^{\pm m}\left(\sum_{\scriptstyle
r=1}^\infty\left[{{1}\over{r-2}}\right]
P_2^{-{r\over2}}P_1^{+{r\over2}}\right)
$$
$$
 \mp 4  m Q_1^{\pm m}\left(\sum_{\scriptstyle
r=1}^\infty\left[{{1}\over{r+2}}\right]
P_2^{+{r\over2}}P_1^{-{r\over2}}\right)+\delta_J^{\pm m}\eqno(37)
$$
$$
{-i Q_2^{\pm m}}_t=\mp {{1}\over{m}}{Q_2^{\pm m}}_{xx} -2
 Q_2^{\pm m}\left(\sum_{\scriptstyle r=1}^\infty Q_2^{-r}Q_2^{+r}\right)
 \pm 4  m Q_2^{\pm m}\left(\sum_{\scriptstyle
r=1}^\infty\left[{{1}\over{r-2}}\right]
P_2^{-{r\over2}}P_1^{+{r\over2}}\right)
$$
$$
 \mp 4  m Q_2^{\pm m}\left(\sum_{\scriptstyle
r=1}^\infty\left[{{1}\over{r+2}}\right]
P_2^{+{r\over2}}P_1^{-{r\over2}}\right)+\delta_L^{\pm m}\eqno(38)
$$

$$
{-i P_1^{\pm {m\over2}}}_t=\mp {{2}\over{m\mp 2}}{P_1^{\pm
{m\over2}}}_{xx}
 \pm 2  (m\mp 1) {P_1^{\pm {m\over2}}}\left(\sum_{\scriptstyle r=1}^\infty
\left[{{1}\over{r-2}}\right] P_2^{-{r\over2}}P_1^{+{r\over2}}\right)
$$
$$
\mp 2(m\pm 1) {P_1^{\pm {m\over2}}}\left(\sum_{\scriptstyle r=1}^\infty
\left[{{1}\over{r+2}}\right] P_2^{+{r\over2}}P_1^{-{r\over2}}\right)
+ P_1^{\pm {m\over2}}\sum_{\scriptstyle r=1}^\infty Q_1^{+r}Q_2^{-r}\eqno(39)
$$
$$
+ P_1^{\pm {m\over2}}\sum_{\scriptstyle r=1}^\infty Q_1^{-r}Q_2^{+r}
\mp  m P_1^{\pm {m\over2}}\sum_{\scriptstyle r=1}^\infty
 Q_2^{+r} Q_2^{-r}+\delta_{G^1}^{\pm {m\over2}}
$$
\newpage
\noindent and
$$
{- P_2^{\pm {m\over2}}}_t=\mp {{2}\over{m\pm 2}}{P_2^{\pm
{m\over2}}}_{xx}
 \pm 2  (m\pm 1) {P_2^{\pm {m\over2}}}\left(\sum_{\scriptstyle r=1}^\infty
\left[{{1}\over{r-2}}\right] P_2^{-{r\over2}}P_1^{+{r\over2}}\right)
$$
$$
\mp 2(m\mp 1) {P_2^{\pm {m\over2}}}\left(\sum_{\scriptstyle
r=1}^\infty \left[{{1}\over{r+2}}\right]
P_2^{+{r\over2}}P_1^{-{r\over2}}\right)
- P_2^{\pm {m\over2}}\sum_{\scriptstyle r=1}^\infty Q_1^{+r}Q_2^{-r}\eqno(40)
$$
$$
- P_2^{\pm {m\over2}}\sum_{\scriptstyle r=1}^\infty Q_1^{-r}Q_2^{+r}
\mp  m P_1^{\pm {m\over2}}\sum_{\scriptstyle r=1}^\infty
 Q_2^{+r} Q_2^{-r}+\delta_{G^2}^{\pm {m\over2}}
$$
\noindent where
$$
 -i\delta_J^{+ m}=
 \sum_{\scriptstyle r=1\atop\scriptstyle r<m}^\infty Q_2^{+r} {Q_{1x}^{+(m-r)}}
-\sum_{\scriptstyle r=1\atop\scriptstyle r<m}^\infty \Big[{{r}\over{m-r}} \Big]Q_1^{+r} {Q_{2x}^{+(m-r)}}
+2\sum_{\scriptstyle r=1\atop\scriptstyle r<2m}^\infty
  \Big[{{r-m}\over{2m-r-2}}\Big]P_2^{+ {r\over2}}
 {P_{1x}^{+(m-{r\over2})}}
$$
$$
-2\sum_{\scriptstyle r=1\atop\scriptstyle
 r<2m}^\infty\Big[{{r-m}\over{2m-r+2}}\Big]P_1^{+ {r\over2}}
{P_{2x}^{+(m-{r\over2})}}
+\sum_{\scriptstyle r=1\atop\scriptstyle
r>m}^\infty Q_2^{+r} {Q_{1x}^{-(r-m)}}
+\sum_{\scriptstyle
r=1\atop\scriptstyle r>m}^\infty \Big[{{r}\over{r-m}} \Big]Q_1^{+r}
{Q_{2x}^{-(r-m)}}
$$
$$
+2\sum_{\scriptstyle r=1\atop\scriptstyle
r>2m}^\infty\Big[{{r-m}\over{r-2m-2}}\Big]P_1^{+ {r\over2}}
{P_{2x}^{-({r\over2}-m)}}
-2\sum_{\scriptstyle r=1\atop\scriptstyle
r>2m}^\infty\Big[{{r-m}\over{r-2m+2}}\Big]P_2^{+{r\over2}}
{P_{1x}^{-({r\over2}-m)}}
+\sum_{\scriptstyle r=1\atop\scriptstyle
r>m}^\infty Q_2^{-(r-m)} {Q_{1x}^{+r}}\eqno(41)
$$
$$
+\sum_{\scriptstyle
r=1\atop\scriptstyle r>m}^\infty \Big[{{(r-m)}\over{r}} \Big]Q_1^{-(r-m)}
{Q_{2x}^{+s}}
+2\sum_{\scriptstyle r=1\atop\scriptstyle
r>2m}^\infty\Big[{{r-m}\over{r+2}}\Big]P_1^{-({r\over2}-m)}
{P_{2x}^{+{s\over2}}}
-2\sum_{\scriptstyle r=1\atop\scriptstyle
r>2m}^\infty\Big[{{r-m}\over{r-2}}\Big]P_2^{-({r\over2}-m)}
{P_{1x}^{+{r\over2}}}
$$
$$
-i \delta_L^{+ m}= -\sum_{\scriptstyle
r=1\atop\scriptstyle r<m}^\infty \Big[{{2r-m}\over{m-r}} \Big]Q_2^{+r}
{Q_{2x}^{+(m-r)}}
-4\sum_{\scriptstyle r=1\atop\scriptstyle
r<2m}^\infty\Big[{{1}\over{2m-r+2}}\Big]P_1^{+ {r\over2}}
{P_{2x}^{+(m-{r\over2})}}
-4\sum_{\scriptstyle r=1\atop\scriptstyle
r<2m}^\infty\Big[{{1}\over{2m-r-2}}\Big]P_2^{+ {r\over2}}
{P_{1x}^{+(m-{r\over2})}}
$$
$$
+\sum_{\scriptstyle
r=1\atop\scriptstyle r>m}^\infty \Big[{{2r-m}\over{r-m}} \Big]Q_2^{+r}
{Q_{2x}^{-(r-m)}}
+4\sum_{\scriptstyle r=1\atop\scriptstyle
r>2m}^\infty\Big[{{1}\over{r-2m+2}}\Big]P_1^{+ {r\over2}}
{P_{2x}^{+({r\over2}-m)}}
+4\sum_{\scriptstyle r=1\atop\scriptstyle
r>2m}^\infty\Big[{{1}\over{r-2m+2}}\Big]P_2^{+ {r\over2}}
{P_{1x}^{-({r\over2}-m)}}\eqno(42)
$$
$$
+\sum_{\scriptstyle
r=1\atop\scriptstyle r>m}^\infty \Big[{{2r+m}\over{r}} \Big]Q_2^{-(r-m)}
{Q_{2x}^{+r}}
-4\sum_{\scriptstyle r=1\atop\scriptstyle
r>2m}^\infty\Big[{{1}\over{r+2}}\Big]P_1^{-({r\over2}-m)}
{P_{2x}^{+{r\over2}}}
-4\sum_{\scriptstyle r=1\atop\scriptstyle
r>2m}^\infty\Big[{{1}\over{r-2}}\Big]P_2^{-({r\over2}-m)}
{P_{1x}^{+{r\over2}}}
$$
$$
 -i\delta_{G^1}^{+ m}= \sum_{\scriptstyle r=1\atop\scriptstyle
2r<m}^\infty\Big[{{1}\over{r}}\Big]P_1^{+( {m\over2}-r)}
{Q_{1x}^{+r}}
 -{{1}\over 2}\sum_{\scriptstyle r=1\atop\scriptstyle
2r<m}^\infty\Big[{{m-3r}\over{r}}\Big]P_1^{+( {m\over2}-r)}
{Q_{2x}^{+r}}
 -2\sum_{\scriptstyle r=1\atop\scriptstyle
2r<m}^\infty\Big[{{1}\over{m-2r-2}}\Big]P_{1x}^{+( {m\over2}-r)}
{Q_{1}^{+r}}
$$
$$
-2\sum_{\scriptstyle r=1\atop\scriptstyle
2r<m}^\infty\Big[{{3r-m}\over{m-2r-2}}\Big]P_{1x}^{+( {m\over2}-r)}
{Q_{2}^{+r}}
 +2\sum_{\scriptstyle r=1\atop\scriptstyle
2r>m}^\infty\Big[{{1}\over{2r-m+2}}\Big]P_{1x}^{-(r-{m\over2})}
{Q_{1}^{+r}}
+2\sum_{\scriptstyle r=1\atop\scriptstyle
2r>m}^\infty\Big[{{3r-m}\over{2r-m+2}}\Big]P_{1x}^{-(r- {m\over2})}
{Q_{2}^{+r}}
$$
$$
 -2\sum_{\scriptstyle r=1\atop\scriptstyle
r>m}^\infty\Big[{{1}\over{r-m}}\Big]P_{1}^{+{r\over2}}
{Q_{1x}^{-({r\over2}-{m\over2})}}
+{1\over2}\sum_{\scriptstyle r=1\atop\scriptstyle
r>m}^\infty\Big[{{3r-m}\over{r-m}}\Big]P_{1}^{+{r\over2}}
{Q_{2x}^{-({r\over2}-{m\over2})}}
 +\sum_{\scriptstyle r=1\atop\scriptstyle
2r>m}^\infty\Big[{{1}\over{r}}\Big]P_{1}^{-(r-{m\over2})}
{Q_{1x}^{+r}}\eqno(43)
$$
$$
+{1\over2}\sum_{\scriptstyle r=1\atop\scriptstyle
2r>m}^\infty\Big[{{3r-m}\over{r}}\Big]P_{1}^{-(r- {m\over2})}
{Q_{2x}^{+r}}
 -2\sum_{\scriptstyle r=1\atop\scriptstyle
r>m}^\infty\Big[{{1}\over{r-2}}\Big]P_{1x}^{+{r\over2}}
{Q_{1}^{-({r\over2}-{m\over2})}}
-{1\over2}\sum_{\scriptstyle r=1\atop\scriptstyle
r>m}^\infty\Big[{{3r-m}\over{r-2}}\Big]P_{1x}^{+{r\over2}}
{Q_{2}^{-({r\over2}-{m\over2})}}
$$
$$
 -i\delta_{G^2}^{+ m}=-\sum_{\scriptstyle r=1\atop\scriptstyle
2r<m}^\infty\Big[{{1}\over{r}}\Big]P_2^{+( {m\over2}-r)}
{Q_{1x}^{+r}}
-{{1}\over 2}\sum_{\scriptstyle r=1\atop\scriptstyle
2r<m}^\infty\Big[{{m-3r}\over{r}}\Big]P_2^{+( {m\over2}-r)}
{Q_{2x}^{+r}}
 -2\sum_{\scriptstyle r=1\atop\scriptstyle
2r<m}^\infty\Big[{{1}\over{m-2r+2}}\Big]P_{2x}^{+( {m\over2}-r)}
{Q_{1}^{+r}}
$$
$$
-2\sum_{\scriptstyle r=1\atop\scriptstyle
2r<m}^\infty\Big[{{3r-m}\over{m-2r+2}}\Big]P_{2x}^{+( {m\over2}-r)}
{Q_{2}^{+r}}
 +2\sum_{\scriptstyle r=1\atop\scriptstyle
2r>m}^\infty\Big[{{1}\over{2r-m-2}}\Big]P_{2x}^{-(r-{m\over2})}
{Q_{1}^{+r}}
+2\sum_{\scriptstyle r=1\atop\scriptstyle
2r>m}^\infty\Big[{{3r-m}\over{2r-m-2}}\Big]P_{2x}^{-(r- {m\over2})}
{Q_{2}^{+r}}
$$
$$
 +2\sum_{\scriptstyle r=1\atop\scriptstyle
r>m}^\infty\Big[{{1}\over{r-m}}\Big]P_{2}^{+{r\over2}}
{Q_{1x}^{-({r\over2}-{m\over2})}}
+{1\over2}\sum_{\scriptstyle r=1\atop\scriptstyle
r>m}^\infty\Big[{{3r-m}\over{r-m}}\Big]P_{2}^{+{r\over2}}
{Q_{2x}^{-({r\over2}-{m\over2})}}
 -\sum_{\scriptstyle r=1\atop\scriptstyle
2r>m}^\infty\Big[{{1}\over{r}}\Big]P_{2}^{-(r-{m\over2})}
{Q_{1x}^{+r}}\eqno(44)
$$
$$
+{1\over2}\sum_{\scriptstyle r=1\atop\scriptstyle
2r>m}^\infty\Big[{{3r-m}\over{r}}\Big]P_{2}^{-(r- {m\over2})}
{Q_{2x}^{+r}}
 +2\sum_{\scriptstyle r=1\atop\scriptstyle
r>m}^\infty\Big[{{1}\over{r+2}}\Big]P_{2x}^{+{r\over2}}
{Q_{1}^{-({r\over2}-{m\over2})}}
+{1\over2}\sum_{\scriptstyle r=1\atop\scriptstyle
r>m}^\infty\Big[{{3r-m}\over{r-2}}\Big]P_{2x}^{+{r\over2}}
{Q_{2}^{-({r\over2}-{m\over2})}}
$$
\noindent and
$$
\delta_J^{-m},\\ \delta_L^{-m},\\ \delta_{G^{1}}^{-m},\\ \delta_{G^{2}}^{-m}
=\\ \delta_{G^{1}}^{+m},\\ \delta_{G^{2}}^{+m},\\ \delta_J^{+m},\\ \delta_L^{+m}\left(%
\begin{array}{c}
  +m\rightarrow-m \\
  +r\rightarrow-r \\
 \end{array}%
\right)\eqno(45)
$$
\section{\bf{AKNS Scheme with N=2 Superconformal Algebra (Ramond Type)}}
\setcounter{equation}{45}
\par We take the soliton  connection as
\begin{equation}
\label{8}
\begin{array}{lll}
\Omega=\Omega_b+\Omega_f
\end{array}
\end{equation}
where
\begin{equation}
\label{8}
\begin{array}{lll}
\Omega_b=
& \Big(&
    i \lambda L_0
+ Q_1^{+m} J_{+m}
+ Q_1^{-m} J_{-m}
+ Q_2^{+m} L_{+m}
+ Q_2^{-m} L_{-m}
 \Big) dx+\\
& \Big(&
-A_1 J_0
-A_2 L_0
+ B_1^{+m} J_{+m}
+ B_1^{-m} J_{-m}
+ B_2^{+m} L_{+m}
+ B_2^{-m} L_{-m}
 \Big) dt%
\end{array}
\end{equation}
\begin{equation}
\label{8}
\begin{array}{lll}
\Omega_f=
& \Big(&
+ P_1^{+{m}} G^1_{+{m}}
+ P_1^{-{m}} G^1_{-{m}}
+ P_2^{+{m}} G^2_{+{m}}
+ P_2^{-{m}} G^2_{-{m}}
 \Big) dx+\\
& \Big(&
-A_3 G^1_0
-A_4 G^2_0
+ C_1^{+{m}} G^1_{+{m}}
+ C_1^{-{m}} G^1_{-{m}}
+ C_2^{+{m}} G^2_{+{m}}
+ C_2^{-{m}} G^2_{-{m}}
 \Big) dt%
\end{array}
\end{equation}

\noindent where $L_0$ , $J_{\pm m}$ , $L_{\pm m}$ are bosonic generators and $G^1_{ {0}}$,$G^2_{ {0}}$,
$G^1_{\pm {m}}$,$G^2_{\pm {m}}$ are fermionic generators of centerless N=2
superconformal algebra of Ramond type , namely they satisfy
the following commutation and anticommutation  relations
\begin{equation}
\label{34}
\begin{array}{lll}
\left[ L_r,L_s\right]&  = & (r-s)\ L_{r+s}\\
\left[ J_r,J_s\right]&  = & 0\\
\left[ L_r,J_s\right]&  = & -s\ J_{r+s}\\
\left\{ G^1_r,G^2_s\right\}&  = & 2\ L_{r+s}+ (r-s)\ J_{r+s}\\
\left[ J_r,G^{1,2}_s\right] & = & \pm G^{1,2}_{r+s}\\
\left[ L_r,G^{1,2}_s\right] & = & ({r\over2}-s)\ G^{1,2}_{r+s}\\
\left\{ G^{1,2}_r,G^{1,2}_s\right\}&  = & 0%
\end{array}
\end{equation}
\noindent  Here, $J_{\pm m}$,$L_{\pm m}$,$G^1_{\pm {m}}$ and $G^2_{\pm {m}}$ are generators with
positive(negative) integer indices . In
Eq.(46) we assume summation over the repeated indices. The fields
$Q_{1,2}^{\pm m}$ and
 $P_{1,2}^{\pm {m}}$are x,t dependent and also functions $A_{1,2}$,$A_{3,4}$,
 $B_{1,2}^{\pm m}$ and $C_{1,2}^{\pm {m}}$ are x,t and $\lambda$ dependent.
 \par From the integrability condition given by Eq.(6) we obtain
$$
A_{1x}= \sum_{\scriptstyle r=1}^\infty
 r ( B_1^{+r} Q_2^{-r}
   - B_1^{-r} Q_2^{+r}
   + B_2^{+r} Q_1^{-r}
   - B_2^{-r} Q_1^{+r}
-2 (
     P_1^{-{r}} C_2^{+{r}}
   - P_1^{+{r}} C_2^{-{r}}
   - P_2^{-{r}} C_1^{+{r}}
   + P_2^{+{r}} C_1^{-{r}}))
   \eqno(50)
$$
$$
A_{2x}=  \sum_{\scriptstyle r=1}^\infty
 (2 r ( B_2^{-r} Q_2^{+r}
   -    B_2^{+r} Q_2^{-r})
   + P_1^{-{r}} C_2^{+{r}}
   + P_1^{+{r}} C_2^{-{r}}
   + P_2^{-{r}} C_1^{+{r}}
   + P_2^{+{r}} C_1^{-{r}})
   \eqno(51)
$$
$$
A_{3x}=-{3\over2} \sum_{\scriptstyle r=1}^\infty
 r ( B_2^{+r} P_1^{-r}
   - B_2^{-r} P_1^{+r}
   + C_1^{+r} Q_2^{-r}
   - C_1^{-r} Q_2^{+r})
- (  B_1^{+{r}} P_1^{-{r}}
   + B_1^{-{r}} P_1^{-{r}}
   - C_1^{+{r}} Q_1^{-{r}}
   - C_1^{-{r}} Q_1^{+{r}}))
   \eqno(52)
$$

$$
A_{4x}=-{3\over2} \sum_{\scriptstyle r=1}^\infty
 r ( B_2^{+r} P_2^{-r}
   - B_2^{-r} P_2^{+r}
   + C_2^{+r} Q_2^{-r}
   - C_2^{-r} Q_2^{+r})
+ (  B_1^{+{r}} P_2^{-{r}}
   + B_1^{-{r}} P_2^{-{r}}
   - C_2^{+{r}} Q_1^{-{r}}
   - C_2^{-{r}} Q_1^{+{r}}))
   \eqno(53)
$$
$$
{Q_1^{\pm m}}_t= {B_1^{\pm m}}_x \mp i m \lambda B_1^{\pm m} \mp m
P_1^{\pm m} A_4 \mp m P_2^{\pm m} A_3\mp m A_2 Q_1^{\pm m} +
\delta_J^{\pm m}
\eqno(54)
$$
$$
{Q_2^{\pm m}}_t= {B_2^{\pm m}}_x \mp i m \lambda B_2^{\pm m}
- 2 P_1^{\pm m} A_4 - 2 P_2^{\pm m} A_3\mp m A_2 Q_2^{\pm m} + \delta_L^{\pm m}
\eqno(55)
$$
$$
{P_1^{\pm m}}_t= {C_1^{\pm m}}_x \mp i m \lambda C_1^{\pm m}
+ P_1^{\pm m}A_1  \mp  P_1^{\pm m} A_2- A_3 Q_1^{\pm m} \mp {{m}}A_3 Q_2^{\pm m}+ \delta_{G_1}^{\pm m}
\eqno(56)
$$
$$
{P_2^{\pm m}}_t= {C_2^{\pm m}}_x \mp i m \lambda C_2^{\pm m}
- P_1^{\pm m}A_1  \mp  P_2^{\pm m} A_2+ A_4 Q_1^{\pm m} \mp {{m}}A_4 Q_2^{\pm m}+ \delta_{G_2}^{\pm m}
\eqno(57)
$$
\noindent where
$$
\delta_J^{+ m}=
\sum_{\scriptstyle r,s=1}^\infty
\left(
 (r-s) P_1^{+r} C_2^{+s}
-(r-s) P_2^{+r} C_1^{+s}
+r Q_1^{+r} B_2^{+s}
-s Q_2^{+r} B_1^{+s}
\right) \delta_{r+s,m}
$$
$$
+\sum_{\scriptstyle r,s=1\atop\scriptstyle
r>s}^\infty
\left(
 (r+s) P_1^{+r} C_2^{-s}
-(r+s) P_2^{+r} C_1^{-s}
+r Q_1^{+r} B_2^{-s}
+s Q_2^{+r} B_1^{-s}
\right) \delta_{r-s,m}\eqno(58)
$$
$$
+\sum_{\scriptstyle r,s=1\atop\scriptstyle
r<s}^\infty
\left(
 (r+s) P_1^{-r} C_2^{+s}
-(r+s) P_2^{-r} C_1^{+s}
+r Q_1^{-r} B_2^{+s}
+s Q_2^{-r} B_1^{+s}
\right) \delta_{-r+s,m}
$$
$$
\delta_L^{+ m}=
\sum_{\scriptstyle r,s=1}^\infty
\left(
 2 P_1^{+r} C_2^{+s}
+2 P_2^{+r} C_1^{+s}
+(r-s) Q_2^{+r} B_2^{+s}
\right) \delta_{r+s,m}
$$
$$
+\sum_{\scriptstyle r,s=1\atop\scriptstyle
r>s}^\infty
\left(
 2 P_1^{+r} C_2^{-s}
+2 P_2^{+r} C_1^{-s}
+(r+s) Q_2^{+r} B_2^{-s}
\right) \delta_{r-s,m}\eqno(59)
$$
$$
+\sum_{\scriptstyle r,s=1\atop\scriptstyle
r<s}^\infty
\left(
 2 P_1^{-r} C_2^{+s}
+2 P_2^{-r} C_1^{+s}
-(r+s) Q_2^{-r} B_2^{+s}
\right) \delta_{-r+s,m}
$$
$$
\delta_{G^1}^{+ m}=
-\sum_{\scriptstyle r,s=1}^\infty
\left(
  B_1^{+s} P_1^{+r}
-{1\over2}(2r-s) B_2^{+s} P_1^{+r}
- C_1^{+s} Q_1^{+r}
-{1\over2}(r-2s) C_1^{+s} Q_2^{+r}
\right) \delta_{r+s,m}
$$
$$
-\sum_{\scriptstyle r,s=1\atop\scriptstyle
r>s}^\infty
\left(
  B_1^{-s} P_1^{+r}
-{1\over2}(2r+s) B_2^{-s} P_1^{+r}
- C_1^{-s} Q_1^{+r}
-{1\over2}(r+2s) C_1^{-s} Q_2^{+r}
\right) \delta_{r-s,m}\eqno(60)
$$
$$
-\sum_{\scriptstyle r,s=1\atop\scriptstyle
r<s}^\infty
\left(
  B_1^{+s} P_1^{-r}
+{1\over2}(2r+s) B_2^{+s} P_1^{-r}
- C_1^{+s} Q_1^{-r}
+{1\over2}(r+2s) C_1^{+s} Q_2^{-r}
\right) \delta_{-r+s,m}
$$
$$
\delta_{G^2}^{+ m}=
+\sum_{\scriptstyle r,s=1}^\infty
\left(
  B_1^{+s} P_2^{+r}
-{1\over2}(2r-s) B_2^{+s} P_2^{+r}
+ C_2^{+s} Q_1^{+r}
-{1\over2}(r-2s) C_2^{+s} Q_2^{+r}
\right) \delta_{r+s,m}
$$
$$
+\sum_{\scriptstyle r,s=1\atop\scriptstyle
r>s}^\infty
\left(
  B_1^{-s} P_2^{+r}
-{1\over2}(2r+s) B_2^{-s} P_2^{+r}
+ C_2^{-s} Q_1^{+r}
-{1\over2}(r+2s) C_2^{-s} Q_2^{+r}
\right) \delta_{r-s,m}\eqno(61)
$$
$$
-\sum_{\scriptstyle r,s=1\atop\scriptstyle
r<s}^\infty
\left(
  B_1^{+s} P_2^{-r}
+{1\over2}(2r+s) B_2^{+s} P_2^{-r}
+ C_2^{+s} Q_1^{-r}
+{1\over2}(r+2s) C_2^{+s} Q_2^{-r}
\right) \delta_{-r+s,m}
$$
\noindent and
$$
\delta_J^{-m},\\ \delta_L^{-m},\\ \delta_{G^{1}}^{-m},\\ \delta_{G^{2}}^{-m}
=\\ \delta_{G^{1}}^{+m},\\ \delta_{G^{2}}^{+m},\\ \delta_J^{+m},\\ \delta_L^{+m}\left(%
\begin{array}{c}
  +m\rightarrow-m \\
  +r\rightarrow-r \\
  +s\rightarrow-s \\
\end{array}%
\right)\eqno(62)
$$
\par In AKNS scheme we expand
$A_1$,$A_2$,$A_3$,$A_4$,$B_1^{\pm m}$,$B_2^{\pm m}$,$C_1^{\pm m}$ and $C_2^{\pm
m}$ in terms of the positive powers of $\lambda$ as
\setcounter{equation}{62}
\begin{equation}
A_1=\sum_{\scriptstyle n=0}^2 \lambda^n a_{1n};\ \
A_2=\sum_{\scriptstyle n=0}^2 \lambda^n a_{2n};\ \
A_3=\sum_{\scriptstyle n=0}^2 \lambda^n a_{3n};\ \
A_4=\sum_{\scriptstyle n=0}^2 \lambda^n a_{4n};\ \
\end{equation}
\begin{equation}
B_1^{\pm m}=\sum_{\scriptstyle n=0}^2 \lambda^n b^{\pm m}_{1n};\ \
B_2^{\pm m}=\sum_{\scriptstyle n=0}^2 \lambda^n b^{\pm m}_{2n};\ \
C_1^{\pm {m}}=\sum_{\scriptstyle n=0}^2 \lambda^n c^{\pm{m}}_{1n};\ \
C_2^{\pm {m}}=\sum_{\scriptstyle n=0}^2 \lambda^n c^{\pm{m}}_{2n}
\end{equation}

\noindent Inserting Eq.(63-64) into Eqs.(50-57)gives 36 relations  in
terms of $a_{1n}$,$a_{2n}$,$a_{3n}$,$a_{4n}$,$b^{\pm m}_{1n}$,$b^{\pm m}_{2n}$,
$c^{\pm m}_{1n}$ and $c^{\pm m}_{2n}$ (n=0,1,2) . By
solving these relations we get
$$
a_{10}= i \sum_{\scriptstyle r=1}^\infty
( Q_1^{+r}Q_2^{-r}
+ Q_1^{-r}Q_2^{+r})
-2i \sum_{\scriptstyle  r=1}^\infty
 (P_2^{-{r}}P_1^{+{r}}
 +P_2^{+{r}}P_1^{-{r}})
;\ a_{11}=a_{12}=0;\ \
$$
$$
a_{20}= 2i \sum_{\scriptstyle r=1}^\infty
( Q_2^{-r}Q_2^{+r})
-2i \sum_{\scriptstyle  r=1}^\infty(
 \Big[{{1}\over{r}}\Big]P_2^{-{r}}P_1^{+{r}}
-
 \Big[ {{1}\over{r}}\Big]P_2^{+{r}}P_1^{-{r}})
;\ a_{21}=a_{22}=-i;\ \
$$
$$
a_{30}= {3\over2}i \sum_{\scriptstyle r=1}^\infty
(P_1^{+r}Q_2^{-r}+P_1^{-r}Q_2^{+r})
-i \sum_{\scriptstyle  r=1}^\infty
 \Big[{{1}\over{r}}\Big]P_1^{+{r}}Q_1^{-{r}}
+i\sum_{\scriptstyle  r=1}^\infty
 \Big[ {{1}\over{r}}\Big]P_1^{-{r}}Q_1^{+{r}}
;\ a_{31}=a_{32}=0;\eqno(65)
$$
$$
a_{40}= {3\over2}i \sum_{\scriptstyle r=1}^\infty
(P_2^{+r}Q_2^{-r}+P_2^{-r}Q_2^{+r})
+i \sum_{\scriptstyle  r=1}^\infty
 \Big[{{1}\over{r}}\Big]P_2^{+{r}}Q_1^{-{r}}
-i\sum_{\scriptstyle  r=1}^\infty
 \Big[ {{1}\over{r}}\Big]P_2^{-{r}}Q_1^{+{r}}
;\ a_{41}=a_{42}=0;\ \
$$
$$
 b^{\pm m}_{10}=\mp {i\over m} {Q_1^{\pm m}}_x ;\
 b^{\pm m}_{11}= Q_1^{\pm m};\
 b^{\pm m}_{12}=0;\
b^{\pm m}_{20}=\mp {i\over m} {Q_2^{\pm m}}_x ;\
b^{\pm m}_{21}= Q_2^{\pm m};\
b^{\pm m}_{22}=0
 $$
$$
c^{\pm {m}}_{10}= \mp {{i}\over {m}}{P_1^{\pm {m} }}_x  ;\
c^{\pm {m}}_{20}= \pm {{i}\over {m}}{P_2^{\mp {m} }}_x  ;\
c^{\pm {m}}_{11}= P_1^{\pm {m}};\
c^{\pm {m}}_{21}= P_2^{\pm {m}};\
c^{\pm {m}}_{12}=c^{\pm {m}}_{22}=0
$$
\noindent By using the relations given by Eq.(65) from Eqs.(54-57)
we obtain the coupled super NLS equations as
$$
{-i Q_1^{\pm m}}_t=
\pm {{1}\over{m}}{Q_1^{\pm m}}_{xx}
\pm \left( \sum_{\scriptstyle r=1}^\infty  P_2^{+r}Q_1^{-r}\right)P_1^{\pm m}
$$
$$
\pm P_2^{\pm m}\left(\sum_{\scriptstyle r=1}^\infty  P_1^{+r}Q_1^{-r}\right)
\pm \left(\sum_{\scriptstyle r=1}^\infty  P_2^{-r}Q_1^{-r}\right)P_1^{\pm m}
\mp P_2^{\pm m}\left(\sum_{\scriptstyle r=1}^\infty  P_1^{-r}Q_1^{+r}\right)
$$
$$
\mp 2 \left( \sum_{\scriptstyle r=1}^\infty  P_2^{-r}P_1^{+r}\right)Q_1^{\pm m}
\pm 2 \left( \sum_{\scriptstyle r=1}^\infty  P_2^{+r}P_1^{-r}\right)Q_1^{\pm m}
\pm 2 m \left( \sum_{\scriptstyle r=1}^\infty  Q_2^{+r}Q_2^{-r}\right)Q_1^{\pm m}\eqno(66)
$$
$$
\mp {3\over2} m \left( \sum_{\scriptstyle r=1}^\infty  P_2^{+r}Q_2^{-r}\right)P_1^{\pm m}
\pm {3\over2} m P_2^{\pm m}\left(\sum_{\scriptstyle r=1}^\infty  P_1^{+r}Q_2^{-r}\right)
$$
$$
\mp {3\over2} m \left( \sum_{\scriptstyle r=1}^\infty  P_2^{-r}Q_2^{+r}\right)P_1^{\pm m}
\pm {3\over2} m P_2^{\pm m}\left(\sum_{\scriptstyle r=1}^\infty  P_1^{-r}Q_2^{+r}\right)
+\delta_J^{\pm m}
$$
$$
{-i Q_2^{\pm m}}_t=
\pm {{1}\over{m}}{Q_2^{\pm m}}_{xx}
-{2\over m} \left( \sum_{\scriptstyle r=1}^\infty  P_2^{+r}Q_1^{-r}\right)P_1^{\pm m}
$$
$$
-{2\over m}P_2^{\pm m} \left( \sum_{\scriptstyle r=1}^\infty  P_1^{+r}Q_1^{-r}\right)
+{2\over m} \left( \sum_{\scriptstyle r=1}^\infty  P_2^{-r}Q_1^{+r}\right)P_1^{\pm m}
+{2\over m}P_2^{\pm m} \left( \sum_{\scriptstyle r=1}^\infty  P_1^{-r}Q_1^{+r}\right)
$$
$$
-3 \left( \sum_{\scriptstyle r=1}^\infty  P_2^{+r}Q_1^{-r}\right)P_1^{\pm m}
+3 P_2^{\pm m} \left( \sum_{\scriptstyle r=1}^\infty  P_1^{+r}Q_2^{-r}\right)\eqno(67)
$$
$$
-3 \left( \sum_{\scriptstyle r=1}^\infty  P_2^{-r}Q_2^{+r}\right)P_1^{\pm m}
+3 P_2^{\pm m}  \left( \sum_{\scriptstyle r=1}^\infty  P_1^{-r}Q_2^{+r}\right)
\mp 2 \left( \sum_{\scriptstyle r=1}^\infty  P_2^{-r}P_1^{+r}\right)Q_2^{\pm m}
$$
$$
\pm 2 \left( \sum_{\scriptstyle r=1}^\infty  P_2^{+r}P_1^{-r}\right)Q_2^{\pm m}
\pm 2 m \left( \sum_{\scriptstyle r=1}^\infty  Q_2^{+r}Q_2^{-r}\right)Q_2^{\pm m}+\delta_L^{\pm m}
$$
$$
{-i P_1^{\pm m}}_t=
\pm {{1}\over{m}}{P_1^{\pm m}}_{xx}
+4 \left( \sum_{\scriptstyle r=1}^\infty  P_2^{+r}P_1^{+r}\right)P_1^{\pm m}
$$
$$
- \left( \sum_{\scriptstyle r=1}^\infty  Q_2^{-r}Q_1^{+r}\right)P_1^{\pm m}
- \left( \sum_{\scriptstyle r=1}^\infty  Q_2^{+r}Q_1^{-r}\right)P_1^{\pm m}
\pm 2 m  \left( \sum_{\scriptstyle r=1}^\infty  Q_2^{+r}Q_2^{-r}\right)P_1^{\pm m}
$$
$$
-{1\over m} \left( \sum_{\scriptstyle r=1}^\infty  Q_1^{-r}P_1^{+r}\right)Q_1^{\pm m}
+{1\over m} \left( \sum_{\scriptstyle r=1}^\infty  Q_1^{+r}P_1^{-r}\right)Q_1^{\pm m}\eqno(68)
$$
$$
+{3\over 2} \left( \sum_{\scriptstyle r=1}^\infty  Q_2^{-r}P_1^{+r}\right)Q_1^{\pm m}
+{3\over 2} \left( \sum_{\scriptstyle r=1}^\infty  Q_2^{+r}P_1^{-r}\right)Q_1^{\pm m}
\mp {1\over 2} \left( \sum_{\scriptstyle r=1}^\infty  Q_1^{-r}P_1^{+r}\right)Q_2^{\pm m}
$$
$$
\pm {1\over 2} \left( \sum_{\scriptstyle r=1}^\infty  Q_1^{+r}P_1^{-r}\right)Q_2^{\pm m}
\pm {3\over 4} m \left( \sum_{\scriptstyle r=1}^\infty  Q_2^{-r}P_1^{+r}\right)Q_2^{\pm m}
\pm {3\over 4} m \left( \sum_{\scriptstyle r=1}^\infty  Q_2^{+r}P_1^{-r}\right)Q_2^{\pm m}+\delta_{G^1}^{\pm m}
$$
$$
{-i P_2^{\pm m}}_t=
\pm {{1}\over{m}}{P_2^{\pm m}}_{xx}
-4 \left( \sum_{\scriptstyle r=1}^\infty  P_2^{-r}P_1^{+r}\right)P_2^{\pm m}
$$
$$
+ \left( \sum_{\scriptstyle r=1}^\infty  Q_2^{-r}Q_1^{+r}\right)P_2^{\pm m}
+ \left( \sum_{\scriptstyle r=1}^\infty  Q_2^{+r}Q_1^{-r}\right)P_2^{\pm m}
\pm 2 m  \left( \sum_{\scriptstyle r=1}^\infty  Q_2^{+r}Q_2^{-r}\right)P_2^{\pm m}
$$
$$
-{1\over m} \left( \sum_{\scriptstyle r=1}^\infty  Q_1^{-r}P_2^{+r}\right)Q_1^{\pm m}
+{1\over m} \left( \sum_{\scriptstyle r=1}^\infty  Q_1^{+r}P_2^{-r}\right)Q_1^{\pm m}\eqno(69)
$$
$$
-{3\over 2} \left( \sum_{\scriptstyle r=1}^\infty  Q_2^{-r}P_2^{+r}\right)Q_1^{\pm m}
-{3\over 2} \left( \sum_{\scriptstyle r=1}^\infty  Q_2^{+r}P_2^{-r}\right)Q_1^{\pm m}
\mp {1\over 2} \left( \sum_{\scriptstyle r=1}^\infty  Q_1^{-r}P_2^{+r}\right)Q_2^{\pm m}
$$
$$
\pm {1\over 2} \left( \sum_{\scriptstyle r=1}^\infty  Q_1^{+r}P_2^{-r}\right)Q_2^{\pm m}
\pm {3\over 4} m \left( \sum_{\scriptstyle r=1}^\infty  Q_2^{-r}P_2^{+r}\right)Q_2^{\pm m}
\pm {3\over 4} m \left( \sum_{\scriptstyle r=1}^\infty  Q_2^{+r}P_2^{-r}\right)Q_2^{\pm m}+\delta_{G^2}^{\pm m}
$$
\noindent where
$$
-i \delta_J^{+ m}=
 \sum_{\scriptstyle r=1}^\infty \Big[{{2r-m}\over{m-r}} \Big]P_1^{+r} {P_{2x}^{+(m-r)}}
-\sum_{\scriptstyle r=1}^\infty \Big[{{r}\over{m-r}} \Big]P_2^{+r} {P_{1x}^{+(m-r)}}
-\sum_{\scriptstyle r=1}^\infty  Q_2^{+r} {Q_{1x}^{+(m-r)}}
$$
$$
+\sum_{\scriptstyle r=1}^\infty \Big[{{r}\over{m-r}} \Big] Q_1^{+r} {Q_{2x}^{+(m-r)}}
-\sum_{\scriptstyle r=1\atop\scriptstyle r>m}^\infty \Big[{{2r-m}\over{r-m}} \Big] P_1^{+r} {P_{2x}^{-(r-m)}}
+\sum_{\scriptstyle r=1\atop\scriptstyle r>m}^\infty \Big[{{2r-m}\over{m-r}} \Big] P_2^{+r} {P_{1x}^{-(r-m)}}
$$
$$
-\sum_{\scriptstyle r=1\atop\scriptstyle r>m}^\infty  Q_2^{+r} {Q_{1x}^{-(r-m)}}
+\sum_{\scriptstyle r=1\atop\scriptstyle r>m}^\infty \Big[{{r}\over{r-m}} \Big] Q_1^{+r} {Q_{2x}^{-(r-m)}}
-\sum_{\scriptstyle r=1\atop\scriptstyle r<m}^\infty \Big[{{2r-m}\over{r}} \Big] P_1^{-(r-m)} {P_{2x}^{+r}}\eqno(70)
$$
$$
+\sum_{\scriptstyle r=1\atop\scriptstyle r<m}^\infty \Big[{{2r-m}\over{r}} \Big] P_2^{-(r-m)} {P_{1x}^{+r}}
-\sum_{\scriptstyle r=1\atop\scriptstyle r<m}^\infty  Q_2^{-(r-m)} {Q_{1x}^{+r}}
-\sum_{\scriptstyle r=1\atop\scriptstyle r<m}^\infty \Big[{{r-m}\over{r}} \Big] Q_1^{+(r-m)} {Q_{2x}^{+r}}
$$
$$
-i \delta_L^{+ m}=
 \sum_{\scriptstyle r=1}^\infty \Big[{{2}\over{m-r}} \Big]P_1^{+r} {P_{2x}^{+(m-r)}}
+\sum_{\scriptstyle r=1}^\infty \Big[{{2}\over{m-r}} \Big]P_2^{+r} {P_{1x}^{+(m-r)}}
-\sum_{\scriptstyle r=1}^\infty \Big[{{2r-m}\over{m-r}} \Big]Q_2^{+r} {Q_{1x}^{+(m-r)}}
$$
$$
+\sum_{\scriptstyle r=1\atop\scriptstyle r>m}^\infty \Big[{{2}\over{r}} \Big]P_1^{+r} {P_{2x}^{-(r-m)}}
+\sum_{\scriptstyle r=1\atop\scriptstyle r>m}^\infty \Big[{{2}\over{r-m}} \Big]P_2^{+r} {P_{1x}^{-(r-m)}}
-\sum_{\scriptstyle r=1\atop\scriptstyle r>m}^\infty \Big[{{2r-m}\over{r-m}} \Big]Q_2^{+r} {Q_{1x}^{-(r-m)}}\eqno(71)
$$
$$
+\sum_{\scriptstyle r=1\atop\scriptstyle r<m}^\infty \Big[{{2}\over{r}} \Big]P_1^{-(r-m)} {P_{2x}^{+r}}
+\sum_{\scriptstyle r=1\atop\scriptstyle r<m}^\infty \Big[{{2}\over{r}} \Big]P_2^{-(r-m)} {P_{1x}^{+r}}
-\sum_{\scriptstyle r=1\atop\scriptstyle r<m}^\infty \Big[{{2r-m}\over{r}} \Big]Q_2^{-(r-m)} {Q_{1x}^{+r}}
$$
$$
-i \delta_{G^1}^{+ m}=
          \sum_{\scriptstyle r=1}^\infty \Big[{{1   }\over{m-r}} \Big]Q_1^{+r} {P_{1x}^{+(m-r)}}
+         \sum_{\scriptstyle r=1}^\infty \Big[{{1   }\over{m-r}} \Big]P_1^{+r} {Q_{1x}^{+(m-r)}}
+{1\over2}\sum_{\scriptstyle r=1}^\infty \Big[{{3r-m}\over{m-r}} \Big]Q_2^{+r} {P_{1x}^{+(m-r)}}
$$
$$
+{1\over2}\sum_{\scriptstyle r=1}^\infty \Big[{{3r-m}\over{m-r}} \Big]P_1^{+r} {Q_{2x}^{+(m-r)}}
-         \sum_{\scriptstyle r=1\atop\scriptstyle r>m}^\infty \Big[{{1    }\over{r-m}}\Big]Q_1^{+r} {P_{1x}^{-(r-m)}}
+         \sum_{\scriptstyle r=1\atop\scriptstyle r>m}^\infty \Big[{{1    }\over{r-m}}\Big]P_1^{+r} {Q_{1x}^{-(r-m)}}
\eqno(72)
$$
$$
-{1\over2}\sum_{\scriptstyle r=1\atop\scriptstyle r>m}^\infty \Big[{{3r-2m}\over{r-m}}\Big]Q_2^{+r} {P_{1x}^{-(r-m)}}
-{1\over2}\sum_{\scriptstyle r=1\atop\scriptstyle r>m}^\infty \Big[{{3r-2m}\over{r-m}}\Big]P_1^{+r} {Q_{2x}^{-(r-m)}}
+         \sum_{\scriptstyle r=1\atop\scriptstyle r<m}^\infty \Big[{{1   }\over{r}}\Big]Q_1^{-(r-m)} {P_{1x}^{+r}}
$$
$$
+         \sum_{\scriptstyle r=1\atop\scriptstyle r<m}^\infty \Big[{{1   }\over{r}}\Big]P_1^{-(r-m)} {Q_{1x}^{+r}}
-{1\over2}\sum_{\scriptstyle r=1\atop\scriptstyle r<m}^\infty \Big[{{3r-m}\over{r}}\Big]Q_2^{-(r-m)} {P_{1x}^{+r}}
-{1\over2}\sum_{\scriptstyle r=1\atop\scriptstyle r<m}^\infty \Big[{{3r-m}\over{r}}\Big]P_1^{-(r-m)} {Q_{2x}^{+r}}
$$
$$
-i \delta_{G^2}^{+ m}=
-         \sum_{\scriptstyle r=1}^\infty \Big[{{1   }\over{m-r}} \Big]Q_1^{+r} {P_{2x}^{+(m-r)}}
+         \sum_{\scriptstyle r=1}^\infty \Big[{{1   }\over{m-r}} \Big]P_1^{+r} {Q_{1x}^{+(m-r)}}
+{1\over2}\sum_{\scriptstyle r=1}^\infty \Big[{{3r-m}\over{m-r}} \Big]Q_2^{+r} {P_{2x}^{+(m-r)}}
$$
$$
+{1\over2}\sum_{\scriptstyle r=1}^\infty \Big[{{3r-m}\over{m-r}} \Big]P_1^{+r} {Q_{2x}^{+(m-r)}}
+         \sum_{\scriptstyle r=1\atop\scriptstyle r>m}^\infty \Big[{{1    }\over{r-m}}\Big]Q_1^{+r} {P_{2x}^{-(r-m)}}
+         \sum_{\scriptstyle r=1\atop\scriptstyle r>m}^\infty \Big[{{1    }\over{r-m}}\Big]P_1^{+r} {Q_{1x}^{-(r-m)}}
\eqno(73)
$$
$$
-{1\over2}\sum_{\scriptstyle r=1\atop\scriptstyle r>m}^\infty \Big[{{3r-2m}\over{r-m}}\Big]Q_2^{+r} {P_{2x}^{-(r-m)}}
-{1\over2}\sum_{\scriptstyle r=1\atop\scriptstyle r>m}^\infty \Big[{{3r-2m}\over{r-m}}\Big]P_1^{+r} {Q_{2x}^{-(r-m)}}
-         \sum_{\scriptstyle r=1\atop\scriptstyle r<m}^\infty \Big[{{1   }\over{r}}\Big]Q_1^{-(r-m)} {P_{2x}^{+r}}
$$
$$
+         \sum_{\scriptstyle r=1\atop\scriptstyle r<m}^\infty \Big[{{1   }\over{r}}\Big]P_1^{-(r-m)} {Q_{1x}^{+r}}
-{1\over2}\sum_{\scriptstyle r=1\atop\scriptstyle r<m}^\infty \Big[{{3r-m}\over{r}}\Big]Q_2^{-(r-m)} {P_{2x}^{+r}}
-{1\over2}\sum_{\scriptstyle r=1\atop\scriptstyle r<m}^\infty \Big[{{3r-m}\over{r}}\Big]P_1^{-(r-m)} {Q_{2x}^{+r}}
$$
\noindent and
$$
\delta_J^{-m},\\ \delta_L^{-m},\\ \delta_{G^{1}}^{-m},\\ \delta_{G^{2}}^{-m}
=\\ \delta_{G^{1}}^{+m},\\ \delta_{G^{2}}^{+m},\\ \delta_J^{+m},\\ \delta_L^{+m}\left(%
\begin{array}{c}
  +m\rightarrow-m \\
  +r\rightarrow-r \\
 \end{array}%
\right)\eqno(74)
$$

\section{\bf{Conclusions}}
\par ~~~~~ Using AKNS scheme and N=2 superconformal algebra of
Neveu-Schwarz and Ramond types we obtain two different new super-
extensions of  coupled Nonlinear Schr\"odinger equations.These
super- extensions of coupled NLS equations  involve infinite number
of bosonic and infinite number of fermionic fields,and
mathematically my work is an extension of the work done in ref.6.
The solution of these NLS equations with infinite number of bosonic
and infinite number of femionic fields is an open problem.


\end{document}